\documentstyle[preprint,aps,epsf,floats]{revtex}

\tighten

\def\bsee{B\rightarrow X_s \ l^+ l^-}
\def\bsgam{B\rightarrow X_s \gamma}

\def\shat{\hat s}

\def\OMIT#1{{}}

\begin{document}

\preprint{\vbox{
\hbox{DOE/ER/41014-10-N97}
}}

\title{Non-$1/m_b^n$ Power Suppressed Contributions to Inclusive $\bsee$ Decays}
\author{Jiunn-Wei Chen, Gautam Rupak and Martin J. 
Savage\footnote{{\tt
savage@phys.washington.edu}}} 
\address{
Department of Physics, University of Washington, \\
Seattle, WA 98195-1560.}

\maketitle

\begin{abstract}

We compute  non-perturbative contributions to $\bsee$ 
that are not explicitly suppressed by 
powers of the $b$-quark mass.
They are proportional to 
$\langle B|\overline{h}\sigma\cdot G h|B\rangle$ 
and arise 
from an interference between the free-quark amplitude and 
higher order terms in the matrix element
of the four-quark operator
$\overline{s}^\alpha\gamma^\mu (1-\gamma_5) c_\alpha
\ \overline{c}^\beta \gamma_\mu (1-\gamma_5)b_\beta$.
This correction 
is found to be small over most of the 
dalitz plot except near the charm threshold.  
Unfortunately, the perturbative computation we have performed 
is invalid near charm threshold 
and we do not except to see the structure found at lowest order
reproduced in the data.
We conclude that these non-perturbative contributions do not significantly modify the 
previous analysis of $\bsee$.
\end{abstract}

\vskip 8.0cm 
\leftline{April 1997}

\vfill\eject

Rare decays of B mesons such as $\bsgam$ and $\bsee$ provide a probe of 
physics at distance scales much smaller than the compton wavelength of the b-quark.
The rates for these processes vanish at tree-level in the standard 
model of electroweak interactions but proceed at one-loop and beyond.
Their GIM ``suppressed'' amplitudes  are dominated by 
the loop contributions from the top quark.   
As such, $\bsgam$ and $\bsee$ are ideal candidates for observing physics 
beyond that of the standard model, e.g. an extended Higgs sector.
However, in order to use them to identify or even claim a hint of new 
physics one must have a robust prediction for decay rates and/or other 
observables in the standard model alone.
Both the inclusive $\bsgam$ (with cuts on the photon energy $E_\gamma$) 
and some exclusive  radiative decay modes such as $B\rightarrow K^*\gamma$
have been observed.
As yet neither inclusive nor exclusive decays $\bsee$ have been observed,
however, the CLEO \cite{cleo94}
upper limit of $1.6\times 10^{-5}$ 
on the 
$\overline{B}^0\rightarrow \overline{K}^{*0} e^+e^-$
branching ratio is within a factor of $\sim 3$
of a standard model estimate.
As theoretical predictions for  exclusive decay widths 
are notoriously difficult to obtain it is the inclusive decays
that one is presently looking toward 
to make a meaningful comparison between theory and experiment.

The heavy quark limit of QCD leads to great simplification in the theoretical analysis
of inclusive decays of B mesons
\cite{incl}.
In particular, the inclusive decays of B mesons can be described using an 
operator-product expansion since the characteristic momentum of the process  
$\sim m_b$ is large compared to the scale of strong interactions 
$\sim \Lambda_{\rm QCD}$.
A systematic expansion of decay observables, such as the rate and differential 
spectra can be performed in  powers of $ \Lambda_{\rm QCD}/m_b$ 
and the strong coupling constant  $\alpha_s (\mu)$.
There are three non-perturbative matrix elements, $\overline{\Lambda}$,
$\lambda_1$ and $\lambda_2$, 
\cite{incllit}\ 
that appear in expressions for such observables
up to order $1/m_b^2$ in the expansion.

The analysis of the inclusive decays $\bsgam$ 
\cite{fls,buv}
and $\bsee$ 
\cite{fls}\ to 
${\cal O}\left(1/m_b^2\right)$
is slightly different to the analogous computations performed for inclusive
semileptonic decays.  
The difference lies in the fact that the 
semileptonic decays proceed at tree-level 
from charged current interactions
while $\bsgam$ and $\bsee$ first occur at one-loop. 
There are contributions to $\bsgam$ and $\bsee$  from 
closed charm quark loops in addition to local operators in the effective theory
resulting from
integrating out the top quark and $W^\pm, Z^0$-bosons
(we do not discuss up-quark contributions).
It was recently suggested that these loop contributions give
power corrections to the rate for $\bsgam$
suppressed by the mass of
the charm quark \cite{vol97}\ as opposed to the mass of the bottom quark
\cite{fls,buv}. 
Explicit computation \cite{vol97}\ showed that these new contributions 
are proportional to the chromomagnetic matrix element 
$\langle B| \overline{h}\sigma\cdot G h|B\rangle$. 
Subsequently, it was shown \cite{lrw,gmnp}\ that such contributions
to $B\rightarrow X_s \gamma$  do not dominate
the $1/m_b^2$ power corrections found in \cite{fls,buv}. 
In addition it was
pointed out \cite{lrw,gmnp}\ that there is an infinite tower of operators
whose contributions are suppressed only by powers of 
$m_b\Lambda_{{\rm QCD}}/m_c^2$ but which appear with 
numerically small coefficients.

The Feynman
diagrams highlighted by Voloshin \cite{vol97}\ yield a term that vanishes
for on-shell photons in addition to the term modifying $\bsgam$. 
In this work we determine the contribution of these diagrams to the 
inclusive decay $\bsee$ where $l$ denotes an electron or a muon. 
We do not re-examine the 
${\cal O}\left(1/m_b^2\right)$ contributions 
\cite{fls,ahhm96}
as we have nothing new to add. 
Further, we do not include the non-renormalization group 
pQCD corrections to the decay width \cite{pQCD}.
Only the leading free-quark amplitude and the loop correction yielding this
new non-perturbative contribution are considered.

For momentum scales much below the mass of the $W^\pm$ gauge boson 
it is appropriate to 
describe these rare decays by an effective hamiltonian
\begin{eqnarray}
{\cal H}_{\rm eff} & = & {4 G_F \over \sqrt{2}} V_{bt} V_{st}^\dagger 
\sum_i\ c_i(\mu)\ {\cal O}_i (\mu)
\ \ \ ,
\end{eqnarray}
where the $c_i(\mu )$ are Wilson coefficients of the operators 
${\cal O}_i(\mu )$ renormalized at a scale $\mu $ 
\cite{gsw89,m93,bm95}. 
We have assumed that $V_{bt} V_{st}^\dagger = -V_{bc} V_{sc}^\dagger $
in writing ${\cal H}_{\rm eff}$ this way and we assume that the leptons
and strange quark
are massless.
In order to fix notation we list the operators 
(we use the same notation as
\cite{gsw89})
\begin{eqnarray}
{\cal O}_1 & = & \overline{s}_{L,\alpha}\gamma^\mu b_{L,\alpha}\ 
\overline{c}_{L,\beta}\gamma_\mu c_{L,\beta}
\nonumber\\
{\cal O}_2 & = & \overline{s}_{L,\alpha}\gamma^\mu b_{L,\beta}\ 
\overline{c}_{L,\beta}\gamma_\mu c_{L,\alpha}
\nonumber\\
{\cal O}_3 & = & \overline{s}_{L,\alpha}\gamma^\mu b_{L,\alpha}\ 
\left[
\overline{u}_{L,\beta}\gamma_\mu u_{L,\beta}\ +\ ....
\overline{b}_{L,\beta}\gamma_\mu b_{L,\beta}\right]
\nonumber\\
{\cal O}_4 & = & \overline{s}_{L,\alpha}\gamma^\mu b_{L,\beta}\ 
\left[
\overline{u}_{L,\beta}\gamma_\mu u_{L,\alpha}\ +\ ....
\overline{b}_{L,\beta}\gamma_\mu b_{L,\alpha}\right]
\nonumber\\
{\cal O}_5 & = & \overline{s}_{L,\alpha}\gamma^\mu b_{L,\alpha}\ 
\left[
\overline{u}_{R,\beta}\gamma_\mu u_{R,\beta}\ +\ ....
\overline{b}_{R,\beta}\gamma_\mu b_{R,\beta}\right]
\nonumber\\
{\cal O}_6 & = & \overline{s}_{L,\alpha}\gamma^\mu b_{L,\beta}\ 
\left[
\overline{u}_{R,\beta}\gamma_\mu u_{R,\alpha}\ +\ ....
\overline{b}_{R,\beta}\gamma_\mu b_{R,\alpha}\right]
\nonumber\\
{\cal O}_7 & = & {e\over 16\pi^2} m_b 
\overline{s}_{L,\alpha}\sigma^{\mu\nu} b_{R,\alpha}\  F_{\mu\nu}
\nonumber\\
{\cal O}_8 & = & {e^2\over 16\pi^2}
\overline{s}_{L,\alpha}\gamma^\mu b_{L,\alpha}\ 
\overline{l}\gamma_\mu l
\nonumber\\
{\cal O}_9 & = & {e^2\over 16\pi^2}
\overline{s}_{L,\alpha}\gamma^\mu b_{L,\alpha}\ 
\overline{l}\gamma_\mu \gamma_5 l
\nonumber\\
{\cal O}_{10} & = & {g\over 16\pi^2} m_b 
\overline{s}_{L,\alpha}\sigma^{\mu\nu} 
T^a_{\alpha\beta} b_{R,\beta} \ G^a_{\mu\nu}
\ \ \ \ .
\end{eqnarray}
Working  at leading order \cite{gsw89}\  we truncate 
the operator basis to 
${\cal O}_1$, ${\cal O}_2$, ${\cal O}_7$, ${\cal O}_8$ and 
${\cal O}_9$.   
The next to leading order results can be found in \cite{m93,bm95}\  
and a complete summary including the non-perturbative and pQCD 
corrections can be found in \cite{ahhm96}.

The rate and other ``short-distance''
observables for the decay arise from an 
operator-product expansion of
\begin{eqnarray}
{\cal T} & = & 
{\rm Im} \left(
i\int\ d^4x\ e^{-i qx}
\langle B|\  T\left[ {\cal H}_{\rm eff} (x) 
{\cal H}^\dagger_{\rm eff} (0) \right]\ 
|B\rangle \right)
\ \ \ ,
\end{eqnarray}
integrated over the appropriate leptonic phase space.
The free quark decay result emerges as the leading term in the OPE of 
${\cal T}$ with  higher order
corrections to the free quark result  arising systematically 
\cite{fls,ahhm96} as a power series in 
$\alpha _s(\mu )$ and $1/m_b^n$. 
The component of ${\cal T}$ previously omitted from 
$\bsgam$ \cite{vol97}\ arises in
\begin{eqnarray}
T\left[c_2 {\cal O}_2 (x) 
\  c_7^* {\cal O}_7^\dagger (0)\ \ +\ \ {\rm h.c.} \right]
\ \ \ \ ,
\end{eqnarray}
where the charm loop closes off via a soft gluon and 
a photon, as shown in figure 1.
The diagrams that modify $\bsee$ are shown in figure 2
(we have not shown the hermitian conjugate).

The amplitude resulting from closing the charm quark legs of ${\cal O}_{2}$
with the emission of a gluon of momentum $k$ with polarization $\eta$, 
and a photon of momentum $p$ with polarization $\lambda$ is
given by (to leading order in $k$)
\begin{eqnarray}
{\cal A} & = & 
{e g_s Q_c\over 4 \pi^2} \ 
\overline{s} T^a \gamma^\mu (1-\gamma_5) b\ 
\left[ 
I_1\ {1\over 2} p_\eta \epsilon^{ab\lambda\mu}p_a k_b
-I_1\ {1\over 2} p\cdot k \epsilon^{a\eta\lambda\mu}p_a
+ I_2\ p^2 \epsilon^{\lambda b\eta\mu} k_b
-I_2\ p_\lambda \epsilon^{ab\eta\mu}p_ak_b
\right]
\ \ \ ,
\end{eqnarray}
where the functions $I_{1,2}$ are given by 
\begin{eqnarray}
I_{1,2}& = & \int_0^1\ dx \int_0^{1-x}\ dy 
{2 x y \ \  , \ \  x (1-x) \over 
\left[ m_c^2 - x(1-x)p^2  - 2 x y p \cdot k\right] }
\ \ \ \ . 
\end{eqnarray}
As $k^2\sim \Lambda _{{\rm QCD}}^2 \ll m_c^2$ we neglected 
it in both the numerator of ${\cal A}$ and denominator of $I_{1,2}$.
However the $p\cdot k$ term appearing in the denominator of 
$I_{1,2}$ is of order
$\sim E_\gamma\ \Lambda _{{\rm QCD}}\sim m_c^2$ and dominates
over the other terms near the charm pair 
threshold ($p^2\sim 4m_c^2$) region.
Before proceeding it is interesting to 
explore the issues raised in \cite{lrw,gmnp}.
We expand the integrals 
$I_{1,2}$ in terms of $p\cdot k$ 
and evalute the coefficients of terms higher order in $\Lambda_{\rm QCD}$.
After performing the angular integration over phase space one finds that
\begin{eqnarray}
<I_1 (s,m_c^2)> & = & 
\sum\limits_{n=even}^\infty\  a_n\ b_n(s)\ 
\left(E_{\gamma}\ \Lambda_{\rm QCD}\right)^{n}
\nonumber\\ 
b_n(s) & = & \int_0^1\ dx
{x^{n+1}(1-x)^{n+2} \over \left(m_c^2- s x(1-x)\right)^{n+1}}
\ \ \ \ ,
\end{eqnarray}
where $E_{\gamma}$ is the energy of the lepton pair in the rest frame of the 
B-meson.
The first few constants $a_n$ are $a_0={1\over 2}$ , $a_2 = {2\over 3}$ ,
$a_4 = {16\over 15}$.
In the case of on-shell photons, $s=p^2=0$ and $E_{\gamma}=m_b/2$, one finds that
the contribution from terms higher order in 
$\Lambda_{\rm QCD}$ converges rapidly. 
When the photon is off its mass-shell as is the case for $\bsee$ and 
the invariant mass of the lepton pair are far from the charm threshold region
the series is again convergent 
(requiring that $m_b-\sqrt{s} \gg\Lambda_{\rm QCD}$).
In contrast,
when the invariant mass of the lepton pair is near the charm threshold region
the series is not convergent.  
This is to be expected since we know that 
the charmonium resonances in this region do not appear at any 
finite order in perturbation theory.
In fact, this 
problem is not isolated to  our computation but is a problem for 
the entire parton level computation of the differential rate.
One expects that suitable smoothed quantities  over this region of 
invariant mass will be reproduced by the same smoothed quantities 
in the parton model by duality.
For the rest of our discussion we will drop the $p\cdot k$ terms so that the integrals 
$I_{1,2}$ become identical
$I_1 = I_2 = I$ where
\begin{eqnarray}
I(s,m_c^2) & = & \int_0^1\ dy\ { y(1-y)^2\over \left[ m_c^2 - y(1-y) s\right]}
\ \ \ .
\end{eqnarray}

The leading contribution to the differential decay rate comes from the 
free quark decay and at leading order in strong interactions \cite{gsw89}\ 
is
\begin{eqnarray}
{d\Gamma^{(0)}\over d\shat} =  
{\alpha^2\over 4\pi^2} {G_F^2 m_b^5\over 192\pi^3} |V_{bt} V_{st}^\dagger|^2
\ (1-\shat)^2 \ 
& & \left[
(1+2 \shat) | c_8(m_b) + g(r,\shat)(c_2(m_b)+3c_1(m_b))|^2
\right.\nonumber\\
& & \left.
\ + \ (1+2 \shat) |c_9(m_b)|^2
\ + \ 4 (1+2/\shat) |c_7(m_b)|^2
\right.\nonumber\\
& & \left.
+ 12 {\rm Re}\left[ c_7(m_b)^* (c_8(m_b) + g(r,\shat)(c_2(m_b)+3c_1(m_b)))\right]
\right]
\ \ \ \ ,
\end{eqnarray}
where $r=m_c/m_b$ is the ratio of quark masses
and $\shat=p^2/m_b^2$ is the invariant mass of the lepton pair.
The function $g(z,\shat)$ arises from closing the charm loop from operators
${\cal O}_{1,2}$ forming a virtual photon, and is given by (at $\mu=m_b$)
\cite{gsw89,m93,bm95}
\begin{eqnarray}
g(z,\shat) & = & -{8\over 9} ln z + {8\over 27} + {4\over 9 }y 
\nonumber\\
&  & - {2\over 9} (2+y)\sqrt{|1-y|}
\left[ \theta(1-y)\left( ln{1+\sqrt{1-y}\over 1-\sqrt{1-y}} + i\pi \right)
+ \theta(y-1) 2 \tan^{-1} {1\over\sqrt{y-1}}
\right]
\ \ \ \ .
\end{eqnarray}
with $y  =  {4z^2\over \shat} $.

Computation of the non-perturbative  term is straightforward but 
tedious and we will not detail it here.
The additional contribution to the differential rate is found to be  
\begin{eqnarray}
{d\Gamma\over d\shat} = & & 
{\alpha^2\over 4\pi^2} \ {G_F^2 m_b^5\over 192\pi^3}\  
|V_{bt} V_{st}^\dagger|^2\ 
{32\lambda_2\over 3 m_b^2}\ 
(1-\shat)^2 \ \nonumber\\
& & 
{\rm Re} \left[ 
\left(  \  (1/\shat + 6- \shat)\  c_7(m_b) 
\ +\ 
(2+\shat)\ (c_8(m_b) \ +\  (c_2(m_b)\ +\ 3\ c_1(m_b)) \ g(r,\shat) \ ) \ 
\right)^*
\right.\nonumber\\
& & \left.\ \ \ \ 
c_2(m_b)\ I(\shat,r)\ 
\right]
\ \ \ \  .
\end{eqnarray}
The nonperturbative matrix element $\lambda_2$ that appears in this 
contribution
is found from the $B^*-B$ mass difference to be
$\lambda_2 \sim 0.1 {\rm GeV^2}$.
Comparing the size of this new non-perturbative 
contribution to the differential decay rate, as shown in figure 3, 
one see that 
it is small except near the charm threshold. 
Integrating the expression for the rates between $s=m_b^2/2$
and $s=m_b^2$ we find the rate to be  decreased by 
$\sim 1.5\%$ due to this new contribution (i.e. smaller than the contribution
from the $1/m_b^2$ operators, $\sim 3\%$). 
Unfortunately near threshold is precisely the region where the above 
computation fails
due to the presense of incalculable long-distance physics 
associated with the  charmonium resonances.
In the region either far above or far below the 
charm threshold the 
correction considered here can be safely neglected.

In conclusion, we have determined a previously omitted 
non-perturbative contribution to $\bsee$, that is not explicitly
suppressed by $1/m_b^n$.
This has the same origin as the $1/m_c^2$ contribution to 
$\bsgam$ recently observed by 
Voloshin\cite{vol97}.
We find that in the regions of lepton invariant mass where
we can reliably compute such contributions they are small.
The computed amplitude has interesting threshold behavior near $s=4 m_c^2$
but unfortunately large 
strong interactions corrections that we cannot determine
will significantly modify the result in this region.

\bigskip\bigskip

\acknowledgements

This work is partially supported by the US Department of Energy.

\begin{figure}
\epsfxsize=10cm
\hfil\epsfbox{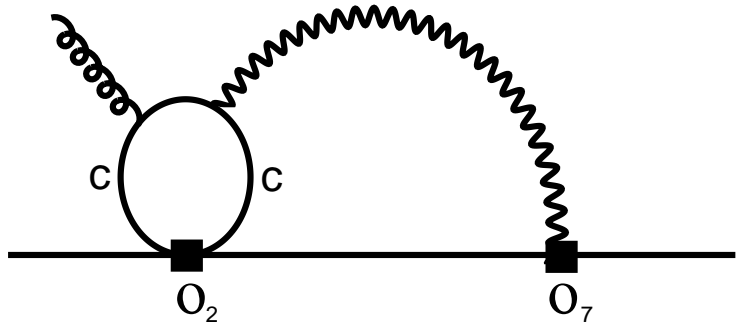}\hfil
\vskip 0.5cm
\caption{A contribution to $\bsgam$ from the time-ordered product 
$T[\  {\cal H}_{\rm eff}(x) \ {\cal H}^{\dagger}_{\rm eff} (0) \ ]$
proportional to $\lambda_2$ and 
naively suppressed by $1/m_c^2$.}
\end{figure}

\begin{figure}
\epsfxsize=14cm
\hfil\epsfbox{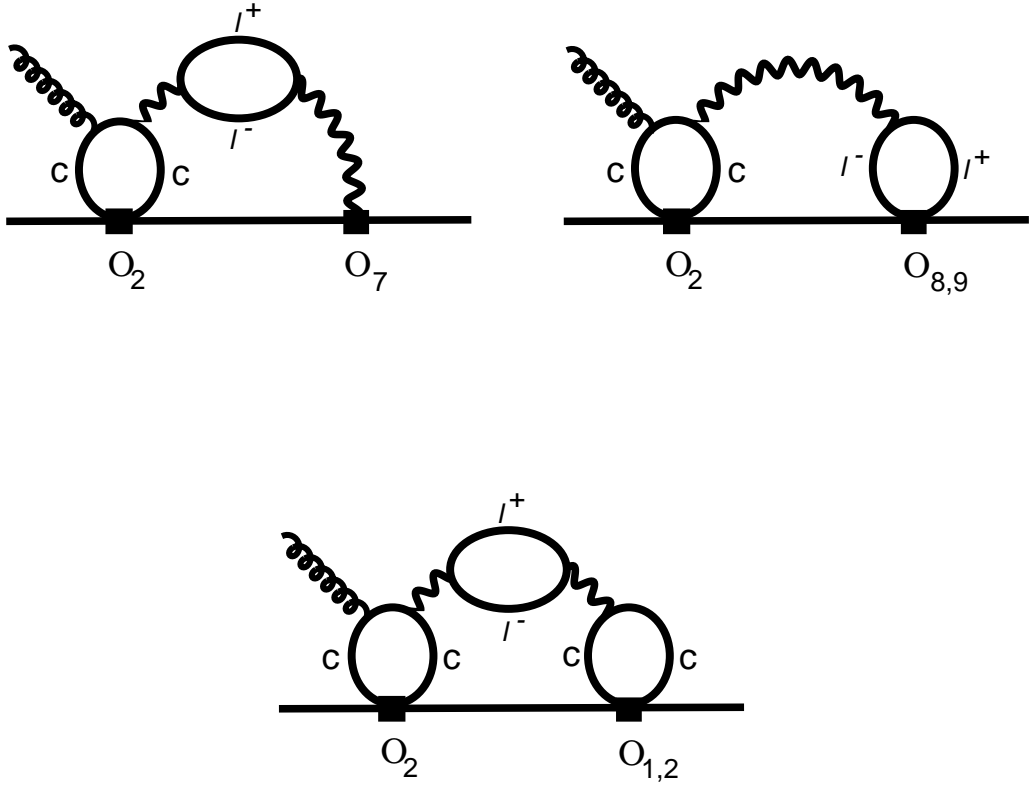}\hfil
\vskip 0.5cm
\caption{Contributions to $\bsee$ from the time-ordered product 
$T[\  {\cal H}_{\rm eff}(x) \ {\cal H}^{\dagger}_{\rm eff} (0) \ ]$
proportional to $\lambda_2$.}
\end{figure}

\begin{figure}
\epsfxsize=14cm
\hfil\epsfbox{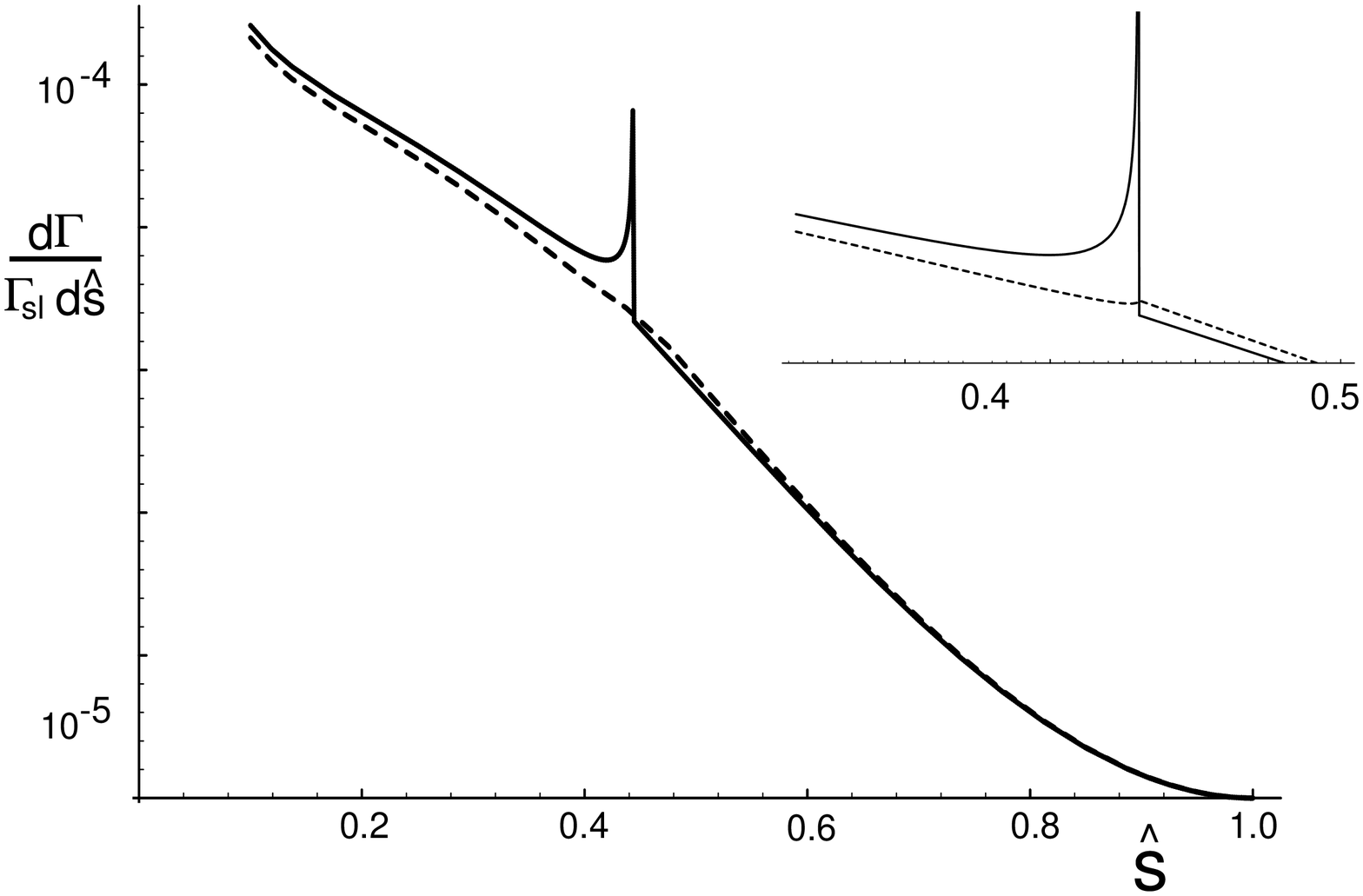}\hfil
\vskip 0.5cm
\caption{The differential decay rate for inclusive $\bsee$ using 
quark masses $m_c = 1.5 \ {\rm GeV}$ and $m_b = 4.5\  {\rm GeV}$
normalized to the semileptonic width.
The dashed line is the leading order, free-quark contribution and 
the solid line is the sum of the free-quark and the non-perturbative
contribution computed in this work.
Inset is the distribution in the charm threshold region.
We have used $c_1=0.22$, $c_2 = -1.09$, $c_7 = 0.31$, $c_8 = -4.16$ and 
$c_9=4.87$ at the scale $\mu=m_b$.}
\end{figure}


\begin{references}

\bibitem{cleo94} R. Balest {\it et al}., (CLEO Collaboration), 
presented at ``International Conference on High Energy Physics'',
Glasgow, Scotland (1994), edited by P.J. Bussey and I.G. Knowles.

\bibitem{incl} M. Voloshin and M. Shifman, Yad. Fiz. 41 (1985) 187; 
J. Chay, H. Georgi and B. Grinstein, Phys. Lett. {\bf B247} (1990) 399.

\bibitem{incllit}I.I. Bigi, N.G. Uraltsev and A.I. Vainshtein, Phys. Lett. {\bf 
B293} (1992) 430; Phys. Lett. {\bf B297} (1993) 477 (E); I.I. Bigi {\rm et
al.}, Phys. Rev. Lett. {\bf 71} (1993) 496;
A.V. Manohar and M.B. Wise, Phys. Rev. {\bf D49} (1994) 1310.

\bibitem{fls}  A. Falk, M. Luke and M.J. Savage, Phys. Rev. {\bf D49} (1994)
3367.

\bibitem{buv}  I.I. Bigi, N.G. Uraltsev and A.I. Vainshtein, Phys. Lett. 
{\bf B247} (1992) 430.

\bibitem{vol97}  M. B. Voloshin, Phys. Lett. {\bf B397} (1997) 275.


\bibitem{lrw}  Z. Ligeti, L. Randall and M.B. Wise, Caltech Preprint
CALT-68-2097 (1997), hep-ph/9702322.

\bibitem{gmnp}  A.K. Grant, A.G. Morgan, S. Nussinov and R.D. Peccei, UCLA
Preprint UCLA/97/TEP/5 (1997).

\bibitem{ahhm96}  A. Ali, G. Hiller, L.T. Handoko and T. Morozumi, 
Phys. Rev. {\bf D55} (1997) 4105.

\bibitem{pQCD} A. Ali and c. Greub, Z. Phys. {\bf C49} (1991) 431; 
Phys. Lett. {\bf B259} (1991) 182;
Z. Phys. {\bf C60} (1993) 433;
Phys. Lett. {\bf B361} (1995) 146;
N. Pott, Phys. Rev {\bf D54} (1996) 938.



\bibitem{gsw89}  B. Grinstein, M.J. Savage and M.B. Wise, Nucl. Phys. {\bf 
B319} (1989) 271.

\bibitem{m93}  M. Misiak, Nucl. Phys. {\bf B393} (1993) 23; Nucl. Phys. {\bf 
B439} (1995) 461 (E);

\bibitem{bm95}  A.J. Buras and M. Munz, Phys. Rev. {\bf D52} (1995) 186.


\end{references}
\end{document}